# Title: Optical Generation of Excitonic Valley Coherence in Monolayer $WSe_2$

Authors: A. M. Jones[1], H. Yu[2], N. J. Ghimire[3,4], S. Wu[1], G. Aivazian[1], J.S. Ross[5], B. Zhao[1], J. Yan[4,6], D. G. Mandrus[3,4,6], D. Xiao[7], W. Yao[2*], X. Xu[1,5*]

**Affiliations:**

[1]Department of Physics, University of Washington, Seattle, Washington 98195, USA

[2]Department of Physics and Center of Theoretical and Computational Physics, University of Hong Kong, Hong Kong, China

[3]Department of Physics and Astronomy, University of Tennessee, Knoxville, Tennessee 37996, USA

[4]Materials Science and Technology Division, Oak Ridge National Laboratory, Oak Ridge, Tennessee, 37831, USA

[5]Department of Materials Science and Engineering, University of Washington, Seattle, Washington 98195, USA

[6]Department of Materials Science and Engineering, University of Tennessee, Knoxville, Tennessee, 37996, USA

[7]Department of Physics, Carnegie Mellon University, Pittsburg, PA 15213, USA

*Correspondence to: wangyao@hku.hk; xuxd@uw.edu

**Abstract:** Due to degeneracies arising from crystal symmetries, it is possible for electron states at band edges ("valleys") to have additional spin-like quantum numbers[1-4]. An important question is whether coherent manipulation can be performed on such valley pseudospins, analogous to that routinely implemented using true spin, in the quest for quantum technologies[5-7]. Here we show for the first time that SU(2) valley coherence can indeed be generated and detected. Using monolayer semiconductor $WSe_2$ devices, we first establish the circularly polarized optical selection rules for addressing individual valley excitons and trions. We then reveal coherence between valley excitons through the observation of linearly polarized luminescence, whose orientation always coincides with that of any linearly polarized excitation. Since excitons in a single valley emit circularly polarized photons, linear polarization can only be generated through recombination of an exciton in a coherent superposition of the two valleys. In contrast, the corresponding photoluminescence from trions is not linearly polarized, consistent with the expectation that the emitted photon polarization is entangled with valley pseudospin. The ability to address coherence[8], in addition to valley polarization[9-13], adds a critical dimension to the quantum manipulation of valley index necessary for coherent valleytronics.

**Main Text**

Monolayer group VI transition metal dichalcogenides (TMDCs) are recently discovered two dimensional (2D) semiconductors[14-16]. They have a direct band-gap in the visible range with the band-edge located at energy degenerate valleys (±K) at the corners of the hexagonal Brillouin zone[17-18]. Initial experiments have demonstrated the formation of highly stable neutral and charged excitons in these monolayer semiconductors, where optically excited electrons and holes are bound together by strong Coulomb interactions[19-20]. In conventional semiconductors, such as GaAs, excitons and trions form at the Brillouin zone center. However, in monolayer TMDCs, confinement of electrons and holes to the ±K valleys gives rise to valley excitons and trions, formed at an energy-degenerate set of non-central points in momentum space.

In principle, these valley excitons offer unprecedented opportunities to dynamically manipulate valley index using optical means, as has been done for optically driven spintronics. Previous work has shown that the structural inversion asymmetry present in monolayer TMDCs gives rise to valley-dependent circularly polarized optical selection rules using the single particle picture[1,9]. Recent observations and electrical control of polarized photoluminescence (PL) in atomically thin molybdenum disulfide are important steps toward the optical generation and detection of valley polarization[9-13]. A more challenging but conceptually appealing possibility is to realize quantum coherence between the two well-separated band extrema in momentum space, i.e. valley quantum coherence, which has not been achieved in solid state systems.

Here we investigate the generation and read out of excitonic intervalley quantum coherence in monolayer $WSe_2$ devices via polarization resolved PL spectroscopy. We obtain monolayer $WSe_2$ through mechanical exfoliation of synthetic $WSe_2$ crystals onto 300 nm $SiO_2$ on heavily doped Si substrates. Figure 1a is a microscope image of a device fabricated by electron beam lithography. We apply PL spectroscopy to investigate the valley excitonic properties in this 2D system. The sample is studied at a temperature of 30 K with an excitation energy of 1.88 eV and spot size of 1.5 μm, unless otherwise specified. We also verified that the applied laser power is within the linear response regime (Figure S1).

Figure 1b shows the 2D spatial map of integrated exciton PL intensity, implying the approximately uniform optical quality of the sample. Figure 1c plots the PL spectrum along the spatial line cut indicated in Fig. 1b, which clearly shows two pronounced excitonic emission features at 1.71 and 1.74 eV. We identified that these two highest energy excitonic emissions are associated with the A exciton[21] by comparing the differential reflection spectrum (black curve) with the PL spectrum[10,19,22] (red curve) in Fig. 1d. These sharp and well separated excitonic features are in clear contrast with the broad spectral width of PL emission in monolayer $MoS_2$[9-11,17-18], but comparable with recent observations in monolayer $MoSe_2$[19].

We assign the exciton species by monitoring the PL emission as a function of gate voltage ($V_g$), which controls the monolayer carrier density. Figure 2a shows the obtained PL map vs. photon energy and $V_g$. The spectral features below 1.675 eV likely arise from phonon side bands which are gate-tunable, but won't be discussed here. When $V_g$ is near zero, this monolayer semiconductor is approximately intrinsic and neutral exciton ($X^o$) emission at 1.749 eV dominates. By either increasing or decreasing $V_g$, excess electrons or holes are injected into the monolayer. $X^o$ tends to capture an extra carrier to form a bounded three-particle system (trion) with smaller emission energy[23]. The data demonstrate that we can either obtain $X^+$ (two holes and one electron) at negative $V_g$ or $X^-$ (two electrons and one hole) at positive $V_g$[19].

At $V_g$ larger than 20 V, a PL feature ($X^{-\prime}$) emerges on the low energy side of $X^-$ and dominates the spectrum as $V_g$ (electron concentration) continues to increase. As a function of $V_g$, the PL intensity, shift in binding energy (inset of Fig. 2b), differential reflection spectrum (Fig. S2), and polarization dependence (presented below) show that $X^{-\prime}$ behaves the same as $X^-$ and thus likely arises from the fine structure of $X^-$. Figure 2b plots the peak intensity of $X^o$, $X^+$, $X^-$, and $X^{-\prime}$ as a function of $V_g$, which clearly shows the gate-tunable excitonic emission.

The binding energy of trions can be directly extracted by taking the energy difference between the trion and neutral exciton in Fig. 2a. Due to the true 2D nature and large effective mass of carriers, the binding energies of trions in monolayer $WSe_2$ are much greater than that in quasi 2D systems[19-20]. Within the applied voltage range, the binding energy of $X^+$ varies from 24 meV near $V_g = 0$ to 30 meV near $V_g = -60$ $V$, while that of $X^-$ changes from 30 meV to 40 meV over a range of 45 V (inset of Fig. 2b). Since the peak position of $X^o$ is nearly independent of $V_g$ while trions redshift, the significant tuning of trion binding energies as a function of $V_g$ is most likely a result of the quantum confined stark effect[24].

We now turn to the investigation of valley exciton polarization. Figure 3a shows the polarization resolved PL spectra at selected gate voltages under $\sigma^+$ light excitation. The complete data set, including all $V_g$ and $\sigma^-$ excitation, can be found in Figs. S3 and S4. The data show that the PL of $X^+$, $X^o$, $X^-$, and $X^{\cdot}$ are all highly circularly polarized. This observation demonstrates that valley optical selection rules[1] derived from the single particle picture are inherited by both neutral and charged excitonic states. It is thus feasible to selectively address valley degrees of freedom through these sharp and well separated exciton species by using circularly polarized optical fields. The ground state configurations for valley excitons and trions and their optical selection rules are schematically shown in Fig. 3b. This optical selectivity differentiates valley excitons from excitons in other systems where polarization is determined by exciton spin configurations.

We also investigate the polarization of valley excitons as a function of photo-excitation energy. Figure 3c shows the polarization resolved PL spectra with $\sigma^+$ excitation for different photon energies. We define the PL polarization as $\rho = \frac{PL(\sigma^+) - PL(\sigma^-)}{PL(\sigma^+) + PL(\sigma^-)}$ where $PL(\sigma^{\pm})$ is the detected PL intensity for polarization $\sigma^{\pm}$. Figure 3d plots the peak polarization of $X^+$ (yellow star), $X^-$ (red circle), $X^o$ (black triangle), and $X^{\cdot}$ (blue square) as a function of photo-excitation energy. The data show that ρ does not decrease significantly as photo-excitation energy increases from 1.79 eV to 2.33 eV, ~80 times the exciton linewidth above the valley exciton emission energy. This remarkable observation is distinct compared with previous valley polarization reported in monolayer $MoS_2$[9-11], where valley polarization is generated only when the photo-excitation energy is within the linewidth of the exciton emission. Our observation suggests that these neutral and charged valley excitons have robust optical selection rules within a large neighborhood of the K point in the Brillouin zone[9], and that interactions during hot-carrier relaxation are not the main mechanisms causing valley depolarization.

The most significant finding is that for linearly polarized light excitation, $X^o$ emission is also highly linearly polarized (ρ=0.4), while trion PL is not. Figure 4a shows the PL spectra at selected $V_g$ under horizontally (H) polarized light excitation at 1.88 eV (see Figs. S5 and S6 for complete data set). The data show that the H component of $X^o$ (black curve) is much stronger than the vertically polarized (V) PL component (red curve), while trions have equal PL intensity under either H or V detection.

Linearly polarized $X^o$ emission has been observed in many other systems such as GaAs quantum wells, quantum dots[25], and carbon nanotubes[26-27]. In these systems, crystal anisotropy is a necessary condition for linear polarization, which predetermines the direction of the polarization axis. However, by investigating the monolayer $WSe_2$ linearly polarized PL for arbitrarily-oriented linearly polarized excitation, we found that the observed $X^o$ polarization is independent of crystal orientation. Figure 4b shows polar plots of $X^o$ peak intensity as a function of detection angle for a given incident linear polarization angle θ. The blue lines are fits using $r = A * (1 + \rho * \cos 2 [x - \varphi])$ where $x$ is the angle of detection, and φ is the $X^o$ polarization angle. We extract φ and plot it as a function of θ in Fig. 4c, which is fit well by a line with a slope of unity. The data demonstrate that φ is completely determined by θ and ρ is isotropic (red squares in Fig. 4c). This implies that the origin of linear polarization in monolayer $WSe_2$ is unique compared to any known solid state system.

We attribute this distinctive finding to the generation of excitonic quantum coherence between opposite valleys by linearly polarized light. We have shown that with $\sigma^+$ ($\sigma^-$) polarized laser excitation $X^o$ is highly polarized at the +K (-K) valley by the circularly polarized valley optical selection rule (Fig. 3b). Since linear polarization is a coherent superposition of $\sigma^+$ and $\sigma^-$, it will simultaneously excite both +K and -K valleys and transfer the optical coherence to valley quantum coherence, i.e. the photo-excited electron-hole pair is a linear superposition in the valley subspace: $\sum_q a_q (\hat{e}^\dagger_{K,q} \hat{h}^\dagger_{K,q} + e^{i2\theta} \hat{e}^\dagger_{-K,q} \hat{h}^\dagger_{-K,q})$, $q$ being the wavevector measured from the K points. The observation of linearly polarized $X^o$ PL parallel to the arbitrarily-oriented linearly polarized excitation implies that the inter-valley coherence has been preserved in the exciton formation.

Key to the preservation of valley quantum coherence is the equivalence of quantum trajectories for the photo-excited electron (hole) on states $|K, q\rangle$ and $|-K, q\rangle$ so that their relative phase remains unchanged in the exciton formation. As the exciton emission is at ~ 1.75 eV, we infer that the 1.88 eV excitation energy is below the electron-hole continuum since the exciton binding energy in monolayer TMDCs is on the order of 0.3-0.5 eV[19,28]. Exciton formation mechanisms include the Coulomb interaction with other carriers and coupling to phonons to transfer the binding energy[29]. The former is dominated by intravalley scattering because of the large momentum space separation between valleys and the long range nature of the Coulomb interaction. The intravalley Coulomb interaction is independent of the valley index and preserves both the valley polarization and valley coherence (Supplementary Materials). Exciton formation through phonon-assisted intravalley scattering also preserves the valley polarization and coherence as such processes are valley independent as well (Fig. 4d). Intervalley scattering through the Coulomb interaction and phonons can result in valley depolarization and valley decoherence (Supplementary Materials). Nevertheless, the required large momentum transfer may render intervalley scattering processes inefficient. Our results could be an indication that intervalley scattering has a comparable or even slower timescale compared to electron-hole recombination, such that valley polarization and valley quantum coherence generated by the excitation laser are manifested in the PL polarization.

The above physical picture is also consistent with the absence of linear polarization for trion emission. For the $X^+$ trion, there are only two possible configurations as shown in Fig. 3b, since the holes at K (-K) only have spin up (down) states due to the giant spin-valley coupling and time reversal symmetry[1]. Upon electron-hole recombination, one $X^+$ configuration becomes a $\sigma^+$ photon ($a^\dagger_{\sigma+}$) plus a spin down hole ($\hat{h}^\dagger_{-K\downarrow}$) and the other becomes a $\sigma^-$ photon ($a^\dagger_{\sigma-}$) plus a spin up hole ($\hat{h}^\dagger_{K\uparrow}$). Their linear superposition can only lead to a spin-photon entanglement: $\hat{h}^\dagger_{K\uparrow} a^\dagger_{\sigma-} + e^{i2\theta} \hat{h}^\dagger_{-K\downarrow} a^\dagger_{\sigma+}$. Linearly polarized photons as a superposition of $a^\dagger_{\sigma-}$ and $a^\dagger_{\sigma+}$ are always forbidden for $X^+$ emission as the hole states associated with $a^\dagger_{\sigma-}$ and $a^\dagger_{\sigma+}$ are orthogonal.

For the $X^-$ trion, although the number of possible configurations is tripled (Fig S8) as electrons at each valley have both spin up and down states, only the linear combination of the two configurations shown in Figure 3b (and the time reversal of this combination) can in principle emit a linearly polarized photon. However, we found that the Coulomb exchange interaction leads to a fine splitting between these two $X^-$ trion configurations. This exchange interaction will likely destroy the valley coherence created by linearly polarized excitation (Supplementary Materials), resulting in the absence of linearly polarized PL for $X^-$. Thus our observation of linearly polarized excitons but not trions strongly indicates optical generation of excitonic valley quantum coherence.

We introduce the Bloch sphere (Fig. 4e) to summarize our work on the optical manipulation of valley pseudo-spins through valley excitons and trions. Circularly polarized light prepares valley pseudo-spin on the north (+K) or south (-K) pole, while arbitrary linearly polarized light prepares states on the equator as a coherent superposition of +K and -K valleys. Since elliptically polarized light would create an imbalance between ±K valleys and likely cause significant valley decoherence, the next challenge is to create arbitrary states on the Bloch sphere away from the poles and equator. We hope that our work will stimulate both experimental and theoretical investigations of carrier scatterings and interactions in monolayer semiconductors, to further reveal the mechanisms and timescales related to valley decoherence and depolarization.

**Acknowledgements:** The authors thank Boris Spivak, David Cobden, Anton Andreev, and Kai-Mei Fu for helpful discussions. This work is mainly supported by the NSF (DMR-1150719). Experimental setup and device fabrication is partially supported by DARPA YFA N66001-11-1-4124. HY and WY were supported by the Research Grant Council of Hong Kong (HKU706412P). NG, JY, DM, and DX were supported by US DoE, BES, Materials Sciences and Engineering Division. Device fabrication was performed at the University of Washington Microfabrication Facility and NSF-funded Nanotech User Facility.

**Author Contributions**

All authors made critical contributions to the work.

**Competing Financial Interests**

The authors declare no competing financial interests.

**Figures**

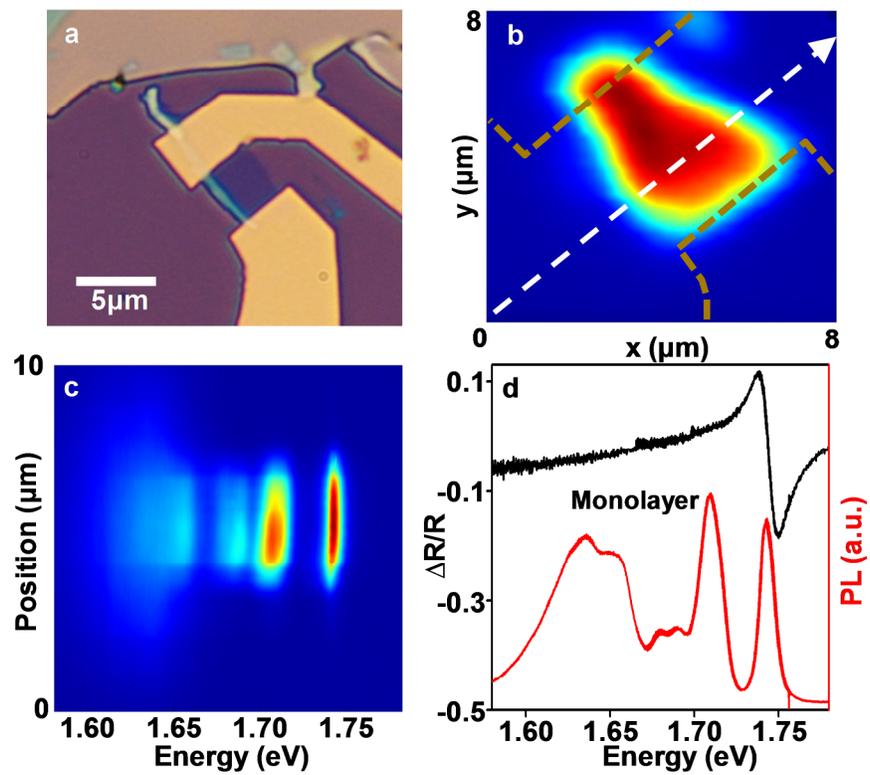

**Figure 1| Device and photoluminescence characterization. a**, Microscope image of a monolayer WSe$_2$ field effect transistor. **b**, Spatial map of integrated exciton photoluminescence of the device in (**a**). **c**, Photoluminescence intensity map as a function of position and photon energy along the line cut in (**b**). **d**, Comparison of differential reflection (black curve) and photoluminescence (red curve) spectra.

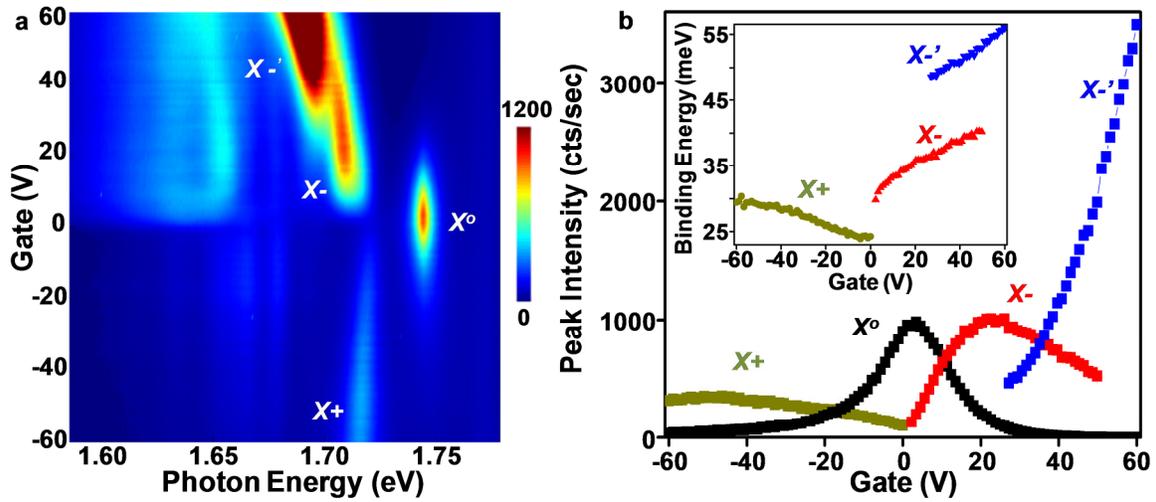

**Figure 2| Electrical control of valley excitons. a**, Photoluminescence intensity map as a function of gate voltage and photon energy. **b**, Peak intensity of exciton and trion photoluminescence as a function of gate voltage. Inset: Trion binding energy as a function of gate voltage.

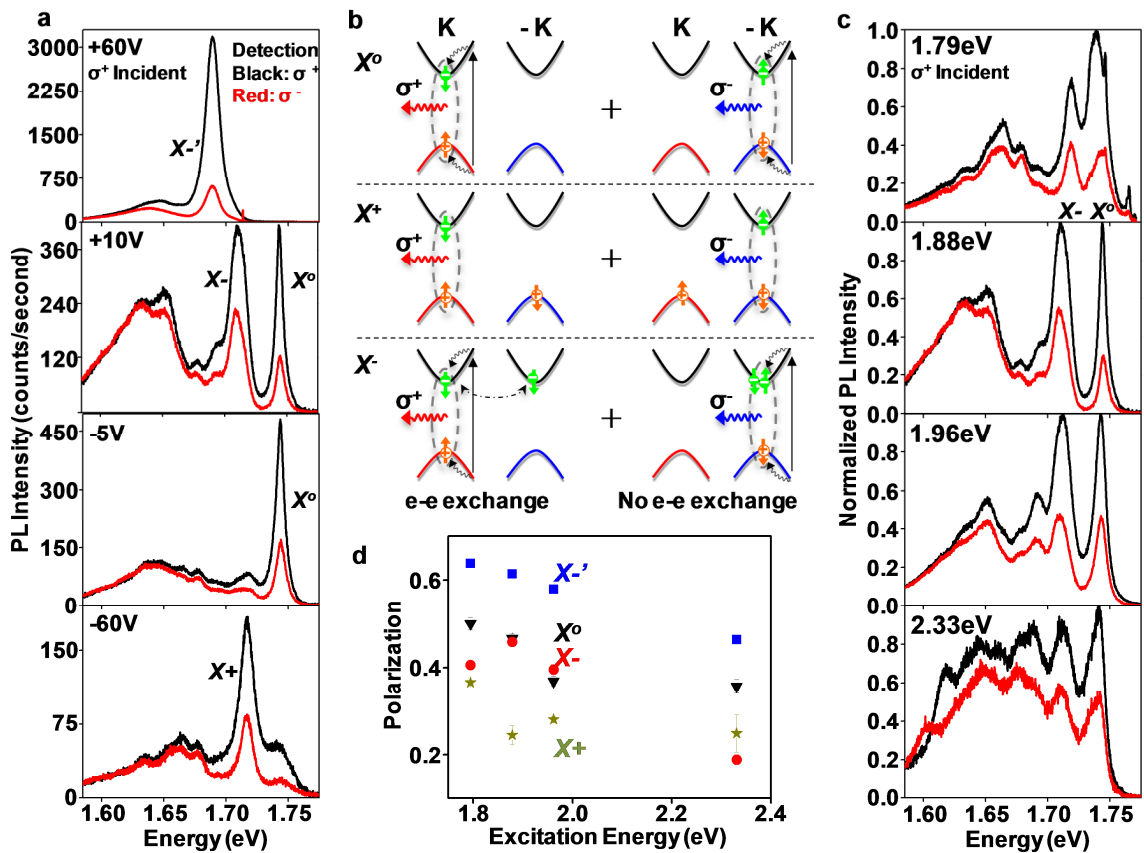

**Figure 3| Optical generation of valley exciton and trion polarization. a**, Polarization resolved photoluminescence spectra at selected gate voltages. Incident laser is $\sigma^+$ polarized. **b**, Cartoons illustrating valley excitons. Both $X^o$ and $X^+$ have two valley configurations, as holes at K (-K) only have spin up (down) states due to the giant spin-valley coupling. The emission of a $\sigma^+$ ($\sigma^-$) photon by $X^+$ leaves behind a spin down (up) hole. $X^-$ has six possible configurations. The two $X^-$ configurations shown here emit, respectively, a $\sigma^+$ and $\sigma^-$ photon leaving a spin down electron in both ±K valleys. Their time reversal configurations (not shown) leave a spin up electron in both ±K valleys upon photon emission. **c**, Polarization resolved photoluminescence spectra as a function of photo-excitation energy (indicated in top left corner). **d**, Degree of valley exciton and trion polarization as a function of photo-excitation energy.

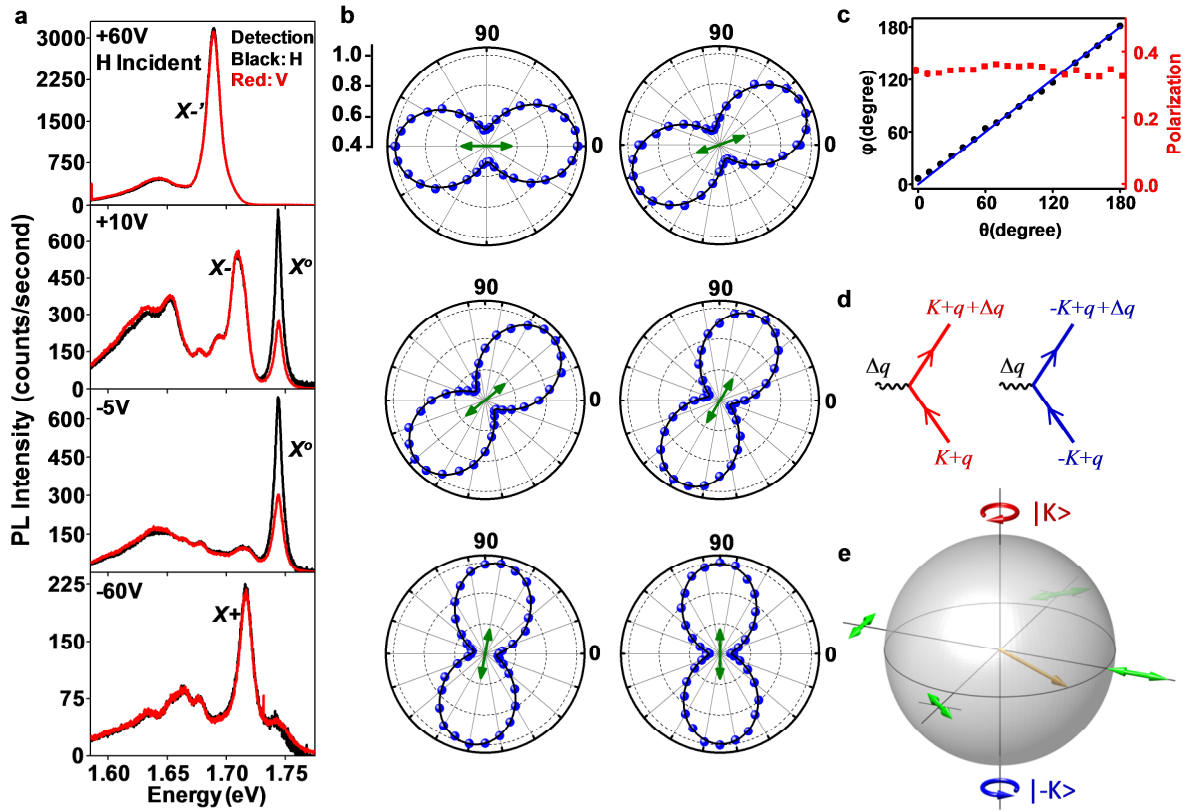

**Figure 4| Signature of excitonic valley quantum coherence. a**, Polarization resolved photoluminescence (PL) spectra at selected gate voltages under horizontally polarized excitation. **b**, Neutral exciton peak intensity as a function of detection angle for given incident laser polarization (marked by green arrow). **c**, Polarization axis φ (black dot) and degree of linear polarization (red square) of neutral exciton PL as a function of incident linear polarization angle θ. **d**, Intravalley scattering Feynman diagram. The wavy line denotes the Fourier component (with wave vector $\Delta q \ll 2K$) of the effective scattering potential by an impurity, phonon, or the Coulomb interaction from other carriers. As the effective scattering potential is identical for valley K and –K, the corresponding quantum trajectories for carriers in states |K,q⟩ and |-K,q⟩ are equivalent, and the relative phase in the valley subspace remains unchanged. **e**, Bloch sphere representation of optical manipulation of excitonic valley pseudo-spins. North and south poles correspond to valley K and –K, respectively, where emitted PL is circularly polarized. State vectors at the equator (orange arrow) are a coherent superposition of the two valleys, corresponding to linearly polarized PL. The PL polarization direction is indicated by arrows outside the sphere.

# Supplementary Materials for

## Optical Generation of Excitonic Valley Coherence in Monolayer WSe$_2$

**Authors:** Aaron M. Jones[1], Hongyi Yu[2], Nirmal J. Ghimire[3,4], Sanfeng Wu[1], Grant Aivazian[1], Jason S. Ross[5], Bo Zhao[1], Jiaqiang Yan[4,6], David G. Mandrus[3,4,6], Di Xiao[7], Wang Yao[2*], Xiaodong Xu[1,5*]

**Affiliations:**
[1] Department of Physics, University of Washington, Seattle, Washington 98195, USA
[2] Department of Physics and Center of Theoretical and Computational Physics, University of Hong Kong, Hong Kong, China
[3] Department of Physics and Astronomy, University of Tennessee, Knoxville, Tennessee 37996, USA
[4] Materials Science and Technology Division, Oak Ridge National Laboratory, Oak Ridge, Tennessee, 37831, USA
[5] Department of Materials Science and Engineering, University of Washington, Seattle, Washington 98195, USA
[6] Department of Materials Science and Engineering, University of Tennessee, Knoxville, Tennessee, 37996, USA
[7] Department of Physics, Carnegie Mellon University, Pittsburg, PA 15213, USA

*Correspondence to: wangyao@hku.hk; xuxd@uw.edu

**This PDF file includes:**

**S1.** Power Dependent Photoluminescence Measurements

**S2.** Gate Dependent Photoluminescence and Differential Reflection Spectrum

**S3.** Gate Dependent Circular and Linear Polarization Studies

**S4.** Valley Depolarization and Decoherence Mechanisms

**S5.** Exchange Splitting in X$^-$

**S6.** Supplementary Figures

**S7.** Supplementary References

## S1. Power Dependent Photoluminescence Measurements

To ensure that the reported PL was within the linear response regime, we examined emitted PL as a function of incident laser power (Fig. S1). Measurements were performed at 30K using 1.88 eV excitation and at gates of +60V for $X^{-'}$, +20V for $X^-$, 0V for $X^o$, and -20V for $X^+$. For each exciton species, a linear dependence of peak PL on laser power is evident for all powers up to 15μW (power used throughout paper). Several other studies of $X^o$ and $X^-$ reveal that the linear power dependence extends beyond 35 μW.

## S2. Gate Dependent Photoluminescence and Differential Reflection Spectrum

In addition to the gate tunable PL presented in the main text, we also obtained the differential reflection spectrum of a broadband white light source as a function of gate, where $\Delta R/R = \frac{R_{SiO_2} - R_{WSe_2}}{R_{SiO_2}}$. The top panel of Figure S2 shows obtained PL (Device 2) as a function of gate and the bottom panel gives the ΔR/R spectrum. The derivative feature associated with the A exciton (cf. Fig. 1D) is strong at near-zero gate voltages. Increasing (decreasing) the gate elicits an increase in differential reflectivity along the same trajectory that $X^-$ ($X^+$) PL follows, while the magnitude of the derivative feature at 1.74 eV decreases. At gate voltages characteristic of the transition from $X^-$ (~20V) to $X^{-'}$ (~40V), ΔR/R is seamless and reasserts the relationship between $X^{-'}$ and $X^-$.

## S3. Gate Dependent Circular and Linear Polarization Studies

Figure S3 (S4) presents the complete set of polarization resolved PL data at each gate voltage studied for right (left) circularly polarized excitation at 1.88 eV. The data show valley exciton and trion polarization at all gates of similar magnitude for both left and right circular incident polarization.

Figure S5 (S6) presents gate dependent linear polarization resolved PL with horizontally (vertically) polarized excitation each gate. We observe that at every gate voltage the linear polarization of $X^o$ is persistent even when trions dominate the spectrum (e.g. $V_g = +/-30V$), while no contribution to linear polarization comes from any trion species.

## S4. Valley Depolarization and Decoherence Mechanisms

For the low energy carriers in the neighborhood of the K or –K points, the direct Coulomb interaction can be separated into intra- and inter-valley parts. Taking the electron-electron interaction as an example, the intra-valley part is written as

$$\hat{V}_{intra} \equiv \Sigma_{\alpha_1,\alpha_2} \Sigma_{\Delta q,q_1,q_2} v(\Delta q) \hat{e}^\dagger_{\alpha_1,q_1-\Delta q} \hat{e}^\dagger_{\alpha_2,q_2+\Delta q} \hat{e}_{\alpha_2,q_2} \hat{e}_{\alpha_1,q_1}.$$

Here $\alpha_{1,2} = K, -K$ is the valley index, $v(\Delta q)$ is the Fourier transform of the Coulomb potential, and Δq is much smaller than the separation between the two valleys. Obviously, $\hat{V}_{intra}$ is independent of the valley index. Thus $\hat{V}_{intra}$ does not affect valley polarization or valley coherence if the two valleys have identical dispersion.

The inter-valley Coulomb interaction is written as

$$\hat{V}_{inter} \equiv v(K) \Sigma_{\Delta q,q_1,q_2} \hat{e}^\dagger_{K,q_1-\Delta q} \hat{e}^\dagger_{-K,q_2+\Delta q} \hat{e}_{K,q_2} \hat{e}_{-K,q_1} + h.c..$$

This interaction will affect the valley polarization and coherence. For example, through $\hat{V}_{inter}$, the mean field $\langle \hat{e}^\dagger_{-K,q_2+\Delta q} \hat{e}_{K,q_2} \rangle$ acts like an effective scattering potential that scatters carriers from one valley to the other, inducing valley depolarization (see Fig. S7).

On the other hand, through $\widehat{V}_{\text{inter}}$, $-\langle e^+_{K,q+\Delta q}\hat{e}_{K,q}\rangle$ ($-\langle e^+_{-K,q+\Delta q}\hat{e}_{-K,q}\rangle$) acts as an effective potential that scatters carrier within valley $-K$ (valley $K$). Thus, the difference $e^+_{K,q+\Delta q}\hat{e}_{K,q} - e^+_{-K,q+\Delta q}\hat{e}_{-K,q}$ is responsible for dynamic change of the relative phase between $|K,q\rangle$ and $|-K,q\rangle$, which leads to valley pure dephasing.

Because of the long range nature of the Coulomb interaction, $v(\Delta q) \gg v(K)$ for $\Delta q \ll K$, thus the valley-independent $\widehat{V}_{\text{intra}}$ will dominate in the exciton formation process assisted by Coulomb scattering with other carriers.

Exciton formation can also be assisted by phonon scattering. For intravalley scattering by long wavelength phonons, the process is also independent of the valley index, i.e. identical trajectories exist for states $|K,q\rangle$ and $|-K,q\rangle$ of a carrier (c.f. Fig. 4d in the main text). Such processes do not affect the valley polarization or the valley coherence. Short wavelength phonons, however, can scatter carriers from one valley to the other (i.e. intervalley scattering), giving rise to valley depolarization.

Scattering by impurities can also play a role in the above exciton formation processes. A smooth impurity scattering potential can only lead to intravalley scattering, but since the process is again valley-independent, it therefore does not affect valley coherence and polarization. Thus, valley depolarization through intervalley scattering is only possible by atomically sharp impurity potentials.

## S5. Exchange Splitting in X⁻

Under linearly polarized excitation, a photo-excited electron-hole pair can be generally written as $\sum_q a_q(\hat{e}^\dagger_{K,q}\hat{h}^\dagger_{K,q} + e^{i2\theta}\hat{e}^\dagger_{-K,q}\hat{h}^\dagger_{-K,q})$. In an n-doped system, the electron-hole pair can then capture a free electron to form a X⁻ trion. As holes in valley K (-K) must be in the spin up (down) state due to the giant spin-valley coupling in the valence band, the ground states of optically bright X⁻ only have six possible configurations as shown schematically in Figure S8. Due to the Pauli exclusion principle the two electrons can either have the same spin but opposite valley index (1st and 4th configurations in Fig S8), opposite spin but same valley index (2nd and 3rd configurations in Fig S8), or opposite spin and valley index (5th and 6th in Fig S8). Only for the 1st and 4th configurations shown can the captured extra electron have an exchange interaction with the excitonic electron and the hole, denoted as $J_{ee}$ and $J'_{eh}$ respectively (see Fig S8). This exchange leads to an energy difference, differentiating these two configurations from the other four. The exchange coupling $J_{eh}$ between the recombining electron and hole (denoted by the dashed ellipse) will also affect the X⁻ energy, but this component is the same for all configurations. We then present $J_{ee}$ and $J'_{eh}$ which are given respectively as (S1):

$$J_{ee} \approx -f_{ee}\int dr_1 dr_2 \psi^*_{Kc\downarrow}(r_1)\psi^*_{-Kc\downarrow}(r_2)v(r_1-r_2)\psi_{Kc\downarrow}(r_2)\psi_{-Kc\downarrow}(r_1)$$
$$\approx -f_{ee}v(K)|\langle u_{-Kc\downarrow}|u_{Kc\downarrow}\rangle|^2,$$
$$J'_{eh} \approx f_{eh}\int dr_1 dr_2 \psi^*_{Kc\downarrow}(r_1)\psi_{-Kv\downarrow}(r_1)v(r_1-r_2)\psi^*_{-Kv\downarrow}(r_2)\psi_{Kc\downarrow}(r_2)$$
$$\approx f_{eh}v(K)|\langle u_{-Kv\downarrow}|u_{Kc\downarrow}\rangle|^2.$$

Here $v(r)$ is the Coulomb potential in real space, and $\psi_{Kc\downarrow}(r) = e^{iKr}u_{Kc\downarrow}(r)$ ($\psi_{-Kv\downarrow}(r) = e^{-iKr}u_{-Kv\downarrow}(r)$) is the conduction (valence) band Bloch state. $f_{ee}$ ($f_{eh}$) corresponds to the modulus square of the envelope function of the electron-electron (electron-hole) relative motion evaluated at zero distance (S2), which is on the order of $1/a_B^2$ where $a_B$ is the Bohr radius of the exciton. From a first principles Bloch function calculation, we find $|\langle u_{-Kc\downarrow}|u_{Kc\downarrow}\rangle|^2 \approx 0.15$ and $|\langle u_{-Kv\downarrow}|u_{Kc\downarrow}\rangle|^2 \approx 0.1$. Thus the exchange-induced splitting between trion configurations is: $J_{ee} + J'_{eh} \sim 0.1 \times v(K)/a_B^2 \sim \frac{0.1}{Ka_B} \times E_b$, where we have taken $v(q) \propto 1/q$ for the Coulomb interaction in 2D and $E_b$ is the exciton binding energy. First principles calculations predict an exciton binding energy $E_b \sim 500$ meV and a Bohr radius $a_B \sim 2$ nm in monolayer MoS$_2$ (S3). In monolayer MoSe$_2$, a trion binding energy of 30 meV has been measured where $E_b \sim 200 - 300$ meV can be inferred (S4). $E_b$ and $a_B$ are expected to take comparable values in monolayer WSe$_2$ due to the comparable bandgap, trion binding energies, and effective masses of electrons and holes. With these numbers, we estimate the X$^-$ configuration splitting to be on the order of 1 meV.

We note that for linearly polarized emission by the X$^-$ trion, only the linear superposition of the 1$^{st}$ and 2$^{nd}$ (or 3$^{rd}$ and 4$^{th}$) configurations is permitted, where the electronic states following the σ+ and σ- photon emission are identical. However, because of the exchange-induced splitting between the 1$^{st}$ and 2$^{nd}$ (3$^{rd}$ and 4$^{th}$) configurations, the valley coherence generated by the linearly polarized laser will be lost upon exciton formation. Thus linearly polarized PL will be absent from the X$^-$ emission.

## S6. Supplementary Figures

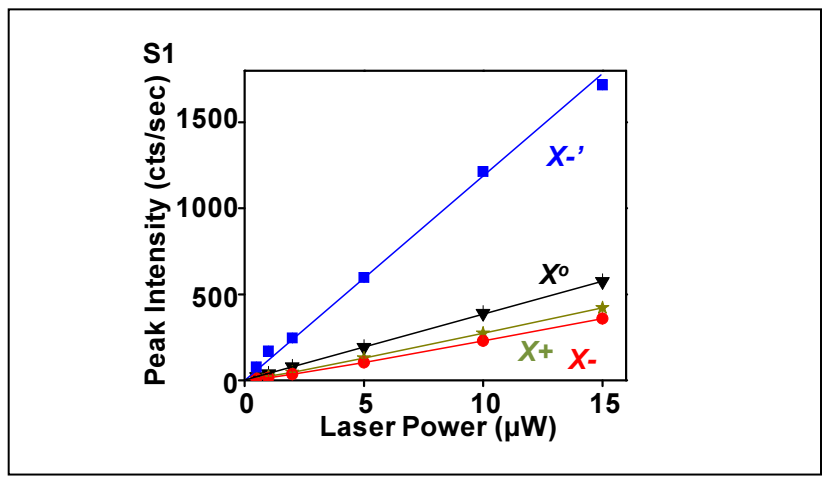

**Figure S1| Photoluminescence power dependence.** Peak photoluminescence intensity as a function of excitation power for $X^{-'}$ (blue square), $X^-$ (red circle), $X^o$ (black triangle), and $X^+$ (yellow star) taken at gate voltages of +60V, +20V, 0V, and -20V, respectively. The linear dependence on power indicates sample response is within the linear regime.

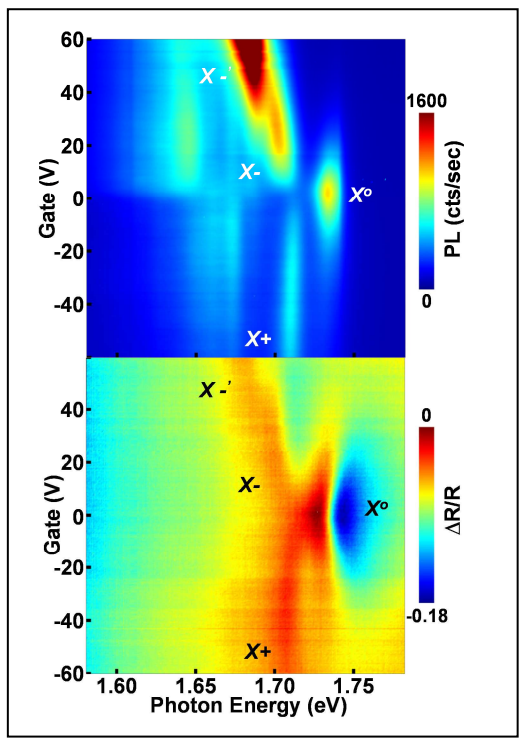

**Figure S2| Gate dependence of photoluminescence and ΔR/R.** PL (top panel) and ΔR/R (bottom panel) vs. gate exhibit the same trajectories associated with each exciton species. The smooth dependence on gate of the X⁻/X⁻' branch indicates a close relationship between the two species.

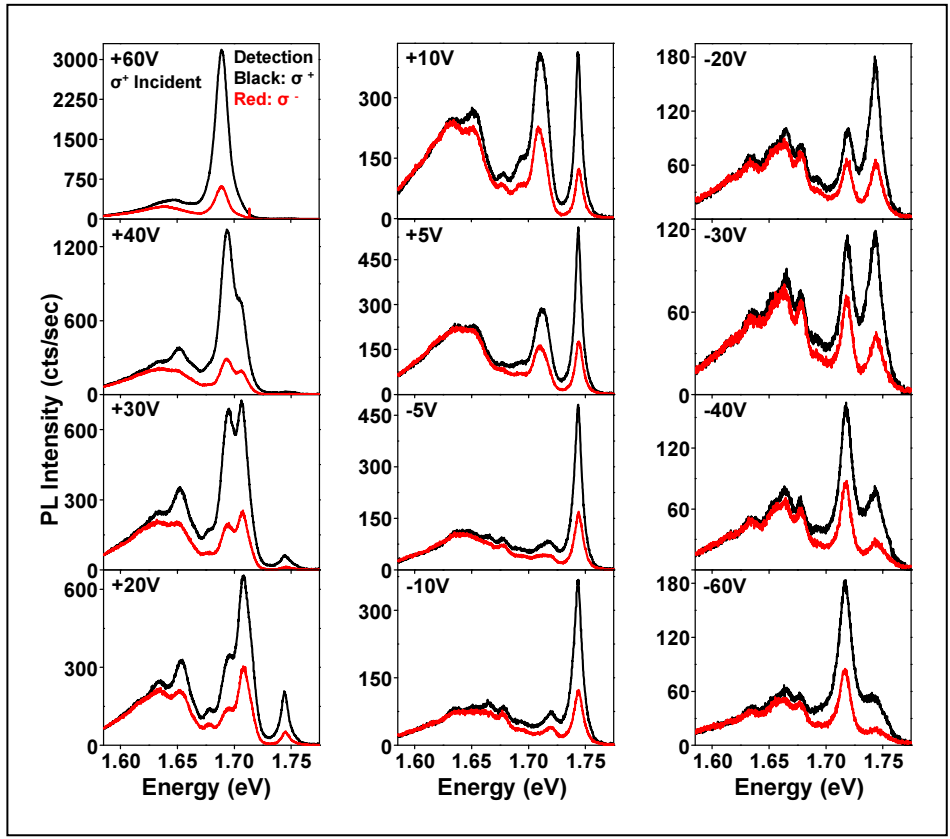

**Figure S3| Right circular polarized excitation.** Polarization dependence as a function of gate (-60V to +60V) for right circular incident polarization and right and left circular polarization detection.

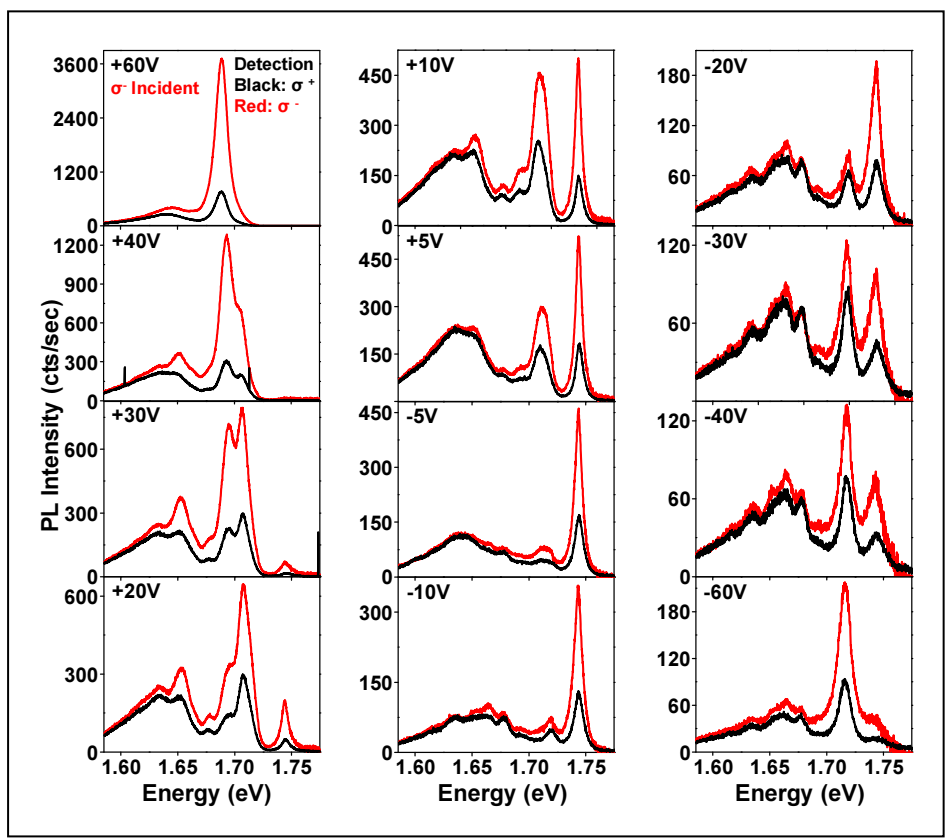

**Figure S4| Left circular polarized excitation.** Polarization dependence as a function of gate (-60V to +60V) for left circular incident polarization and left and right circular polarization detection.

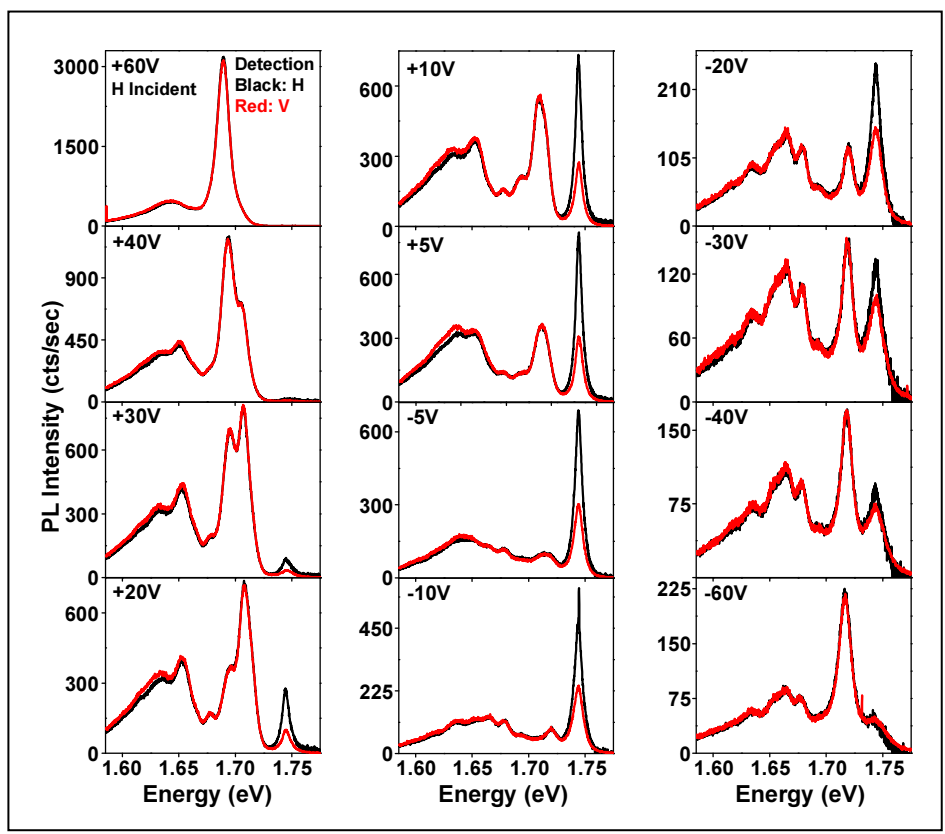

**Figure S5| Horizontally polarized excitation.** Polarization dependence as a function of gate (-60V to +60V) for horizontal incident polarization and horizontal and vertical polarization detection. Over the tunable gate range, only the $X^o$ exciton species exhibits linear polarization.

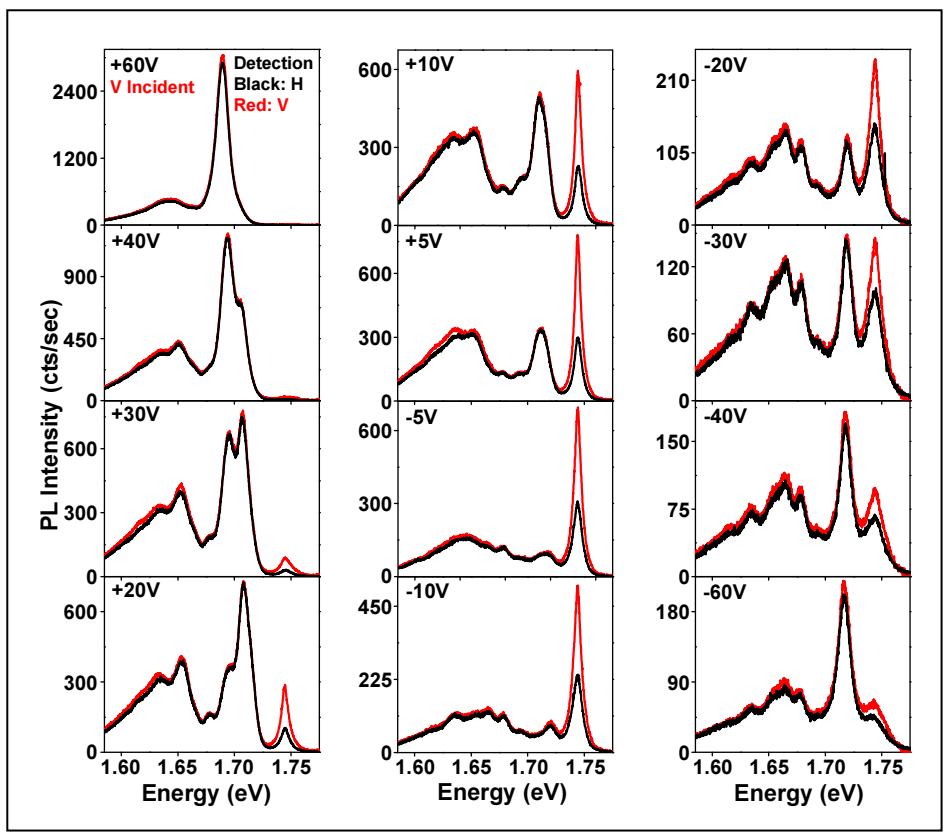

**Figure S6| Vertically polarized excitation.** Polarization dependence as a function of gate (-60V to +60V) for vertical incident polarization and vertical and horizontal polarization detection. Over the tunable gate range, only the $X^o$ exciton species exhibits linear polarization.

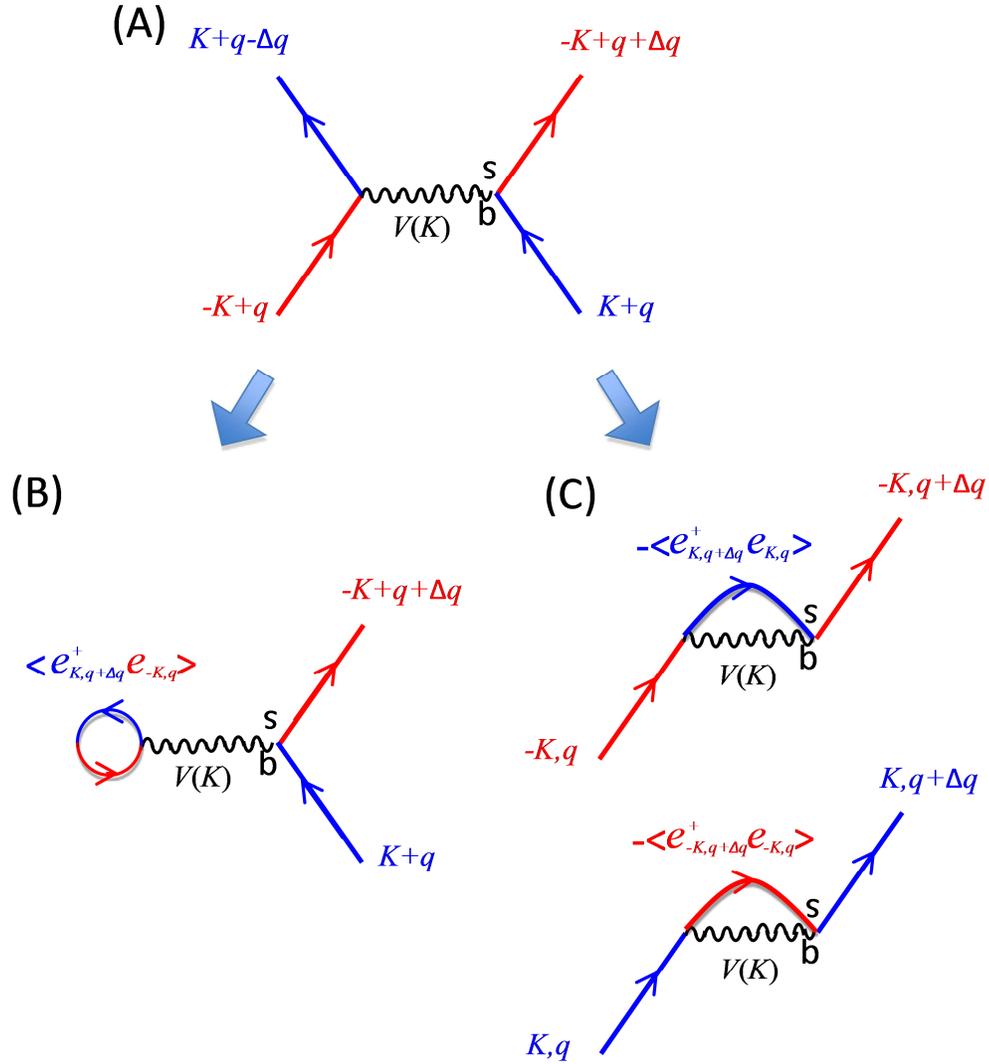

**Figure S7| a**, Intervalley Coulomb scattering. The wavy line denotes the Fourier component (with wavevector ~$K$) of the Coulomb potential. **b**, $\langle e^+_{K,q+\Delta q} \hat{e}_{-K,q} \rangle$ acts as an effective potential that scatters carriers from one valley to another. **c**, $-\langle e^+_{K,q+\Delta q} \hat{e}_{K,q} \rangle$ ($-\langle e^+_{-K,q+\Delta q} \hat{e}_{-K,q} \rangle$) acts as an effective potential that scatters carriers within valley $-K$ (valley $K$). The difference $e^+_{K,q+\Delta q} \hat{e}_{K,q} - e^+_{-K,q+\Delta q} \hat{e}_{-K,q}$ can therefore lead to valley pure dephasing.
.

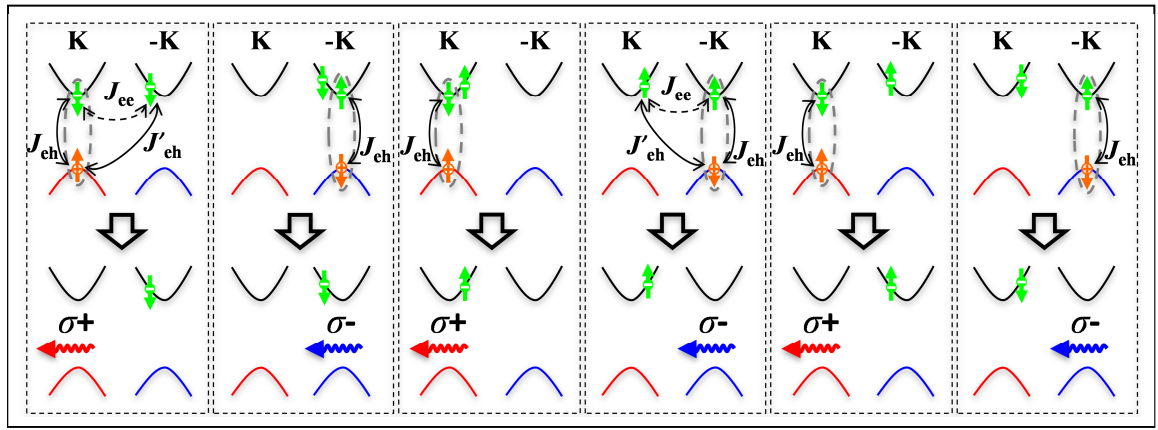

**Figure S8| Exchange Interaction in X$^-$.** Top row: the six possible configurations of the X$^-$ trion. Exchange couplings between the electron and hole components are illustrated. Bottom row: the emitted photon and electron left behind upon electron-hole recombination.

## S7. Supplementary Reference